\documentclass[12pt,manuscript]{aastex}
\usepackage{color}

\slugcomment{}

\shorttitle{8 Planets in 4 Multi-planet Systems}
\shortauthors{Yang M. et al.}

\begin{document}

\title{8 Planets in 4 Multi-planet Systems via TTVs in 1350 Days}

\author{Yang Ming, Liu Hui-Gen, Zhang Hui, Yang Jia-Yi, Zhou Ji-Lin}

\affil{Department of Astronomy \& Key Laboratory of Modern Astronomy and Astrophysics in Ministry of Education, Nanjing University, Nanjing 210093, China}
\email{zhoujl@nju.edu.cn,huigen@nju.edu.cn}

\begin{abstract}
	Analysis of the transit timing variations (TTVs) of candidate pairs near mean-motion resonances (MMRs) is an effective method to confirm planets. Hitherto, 68 planets in 34 multi-planet systems have been confirmed via TTVs. We analyze the TTVs of all candidates from the most recent {\it Kepler} data with a time span as long as about 1350 days (Q0-Q15). The anti-correlations {\bf of TTV signals} and the mass upper limits of candidate pairs in the same system are calculated, using an improved method suitable for long-period TTVs. If the false alarm probability (FAP) of a candidate pair is less than $10^{-3}$ and the mass upper limit for each candidate is less than 13 ${\rm M}_{\rm J}$, we confirm them as planets in the same system. Finally, 8 planets in 4 multi-planet systems are confirmed via analysis of their TTVs. All of the 4 planet pairs are near first-order MMRs, including KOI-2672 near 2:1 MMR, KOI-1236, KOI-1563 and KOI-2038 near 3:2 MMR. Four planets have relatively long {\bf orbital} periods ($>$ 35 day). KOI-2672.01 has a period of 88.51658 {\bf days} and a fit mass {\bf of} $17\,{\rm M}_\oplus$. It is the longest-period planet confirmed, to date, near first-order {\bf MMR} via TTVs.
\end{abstract}


\section{INTRODUCTION}

{\it Kepler} is a landmark space telescope designed to search for Earth-sized planets in and near the habitable zone of Sun-like stars \citep{borucki10}. A transit signature can be detected by {\it Kepler} owning to its high photometric precision of $20$ ppm \citep{koch10}. {\it Kepler} has made significant achievements since it was launched on March 6, 2009. 1235 candidates were identified using the first four months of data \citep{borucki11} and it {\bf numbers over} 2300 as the observation time increased to 16 months \citep{Batalha13}. By now $138$ planets have been confirmed by {\it Kepler}. Recently a new set of data has been released (Q0-Q15 from MAST, http://archive.stsci.edu/kepler). The observation duration of {\it Kepler} is extended to $\sim 1,350$ day. {\bf The number of candidates has} reached as many as $3548$, including $1475$ candidates in multiple systems (up to July, 2013). Due to longer observation time, more long-period candidates are found, e.g. KOI-3946.01 with a period of 308.545 day. Some single transiting systems are {\bf seen} to be multiple systems when more transiting candidates are found, e.g. KOI-255\footnote{\small http://exoplanetarchive.ipac.caltech.edu}. 

For multi-planet systems, the interactions between planets will cause variations of their transit midtimes \citep{agol05,hm05}. Kepler-9 was confirmed as a multi-planet system via transit timing variations (TTVs) successfully by \citet{holman10}. Even non-transiting planets can be detected via TTVs \citep{ballard11,nesvorn12}. TTVs for a two-planet system will generally be anti-correlated because of conservation of energy. By recognizing such anti-correlation and computing the amplitudes of their TTVs, one can infer the mass upper limits of these candidates and therefore confirm them as planets \citep{ttvs7,xie12}. A series of papers have described the theory in detail by the Kepler group {\bf \citep{ttvs2,ttvs3,ttvs4,ttvs7}}. They have confirmed 48 planets in 23 systems combining the anti-correlation method and the dynamical simulation of orbital stability. \citet{xie12} also confirmed 12 planet pairs via TTVs. The masses of 3 planet pairs are exactly calculated while the mass upper limits of others are inferred via the amplitudes of their TTVs.

Hitherto, all the confirmed planet pairs via TTVs are with an observational time span less than $900$ days \citep{ttvs7,carter12}. Although some planets with long periods can be confirmed via TTVs, e.g. Kepler-30 d with a period about $143.2$ day \citep{ttvs4}, confirmations of these planets are less convincing and their masses are quite uncertain due to limited number of observed transits. In this paper, we utilize the recent data released on July 1st, 2013 (Q0-Q15). With a time span as long as 1350 day, we can search for planet pairs near the first order MMRs with long-period TTVs. Their mass upper limits can be inferred via the amplitude described by \citet{lithwick12}.

The arrangement of this paper is as follows. We present how we obtain TTVs of {\bf planet} candidates and describe the confirmation method in Section 2,  i.e. recognizing anti-correlations, calculating false alarm probabilities (FAPs) and estimating mass limits. The properties of the 8 confirmed planets in 4 systems are listed and compared with planets confirmed via TTVs previously in section 3. Furthermore, we show the Gaussian-fit masses in section 4 and compare them with current planet systems in section 5. We summarize and discuss our results in section 6.

\section{TTV ANALYSIS METHOD}

We use the {\it Kepler} data in long cadence (LC) to calculate TTVs. We analyze all {\it Kepler} objects of interest (KOIs) flagged as ``CANDIDATE". We can obtain the corresponding orbital parameters and stellar properties from NASA Exoplanet Archive \citep[http://exoplanetarchive.ipac.caltech.edu;][]{akeson13}.

We compute TTVs for all candidates following the steps described by \citet{xie12}. But some improvements are made. When we calculate TTVs, we cut the light curves into small segments. Each segment centers around a transit midtime and has a length of four times of the transit duration. For KOIs with periods smaller than one month, we eliminate segments contaminated by other candidate(s) to obtain accurate observed midtimes. For KOIs with periods longer than one month, the segments are inevitably easy to be contaminated due to their long transit durations. However, if transits of different candidates in one contaminated segment are not overlapped with a distinguishable separation, we still estimate the transit midtime for each candidate. {\bf We detrend the wings of every segment with first to third-order polynomials}. The fit with the lowest chi-square is adopted. Since {\it Kepler} data have a high precession of $20$ ppm, we can verify {\bf TTV signals} with amplitudes of several minutes.

To confirm planet pairs in a multiple system, two criterions must be satisfied. Firstly they must be anti-correlated so as to infer that the two candidates are in the same system. The anti-correlation can exclude the conditions that the transits are caused by some external reasons, e.g. background eclipsing binaries or background stars transited by planets, etc. To estimate the probability that the observed anti-correlation is caused by random fluctuations, we also check the false alarm probabilities (FAPs) for candidate pairs. Only with a low FAP of less than $10^{-3}$, the candidate pairs are considered to interact with each other reliably. Secondly the physical parameters of the candidate pairs must be in acceptable ranges, especially their masses must be less than 13 $M_{\rm J}$ \citep{spiegel11}. We are interested in candidate pairs near first-order MMRs, namely, the period ratio $P_2/P_1$ for each pair is around $j:j-1$. $P_1$ and $P_2$ are the periods of the inner and outer candidate. We use a normalized distance to resonance parameter $\Delta$ to select out such candidate pairs. $\Delta$ is defined as
\begin{equation} \label{eq_delta}
\Delta = \frac{P_2}{P_1} \frac{j-1}{j} - 1.
\end{equation}
All confirmed planets via TTVs near first-order MMRs satisfy $|\Delta|<0.06$ as shown in Figure \ref{mmr_hist}(b). We adopt this threshold to select candidate pairs. We check the anti-correlated periods of the adopted candidate pairs, calculate their FAPs and set a threshold of $10^{-3}$, and estimate their mass upper limits via TTV amplitudes. These steps will be described in detail in the following subsections.
	
\subsection{ TTV Anti-correlations and FAPs}
	
TTV anti-correlation can indicate the perturbations between two planets in the same system because of the conservation of energy. Transit midtimes of two planets will vary simultaneously and oppositely. Many multiple systems show such characteristics. The anti-correlation method has been  described firstly by \citet{ttvs3}. We modify it by considering a linear trend mixed in long-period TTVs. When computing TTVs a simple linear fitting of transit midtimes has been applied. However it will not completely remove the secular trend of long-period TTVs. Although an untreated linear trend in the TTV signal does not affect the anti-correlation tendency, it will affect TTV amplitude and cause an inaccurate estimation of mass upper limit. Thus we fit TTVs for each KOI using the following function:
\begin{equation}
f = A {\rm{sin}} \left( \frac{2 \pi t}{P} \right) +  B {\rm{cos}} \left( \frac{2 \pi t}{P} \right) + C t + D \label{eq_f}
\end{equation}
where $A$, $B$, $C$ and $D$ are model parameters with test period $P$. We vary $P$ with a wide range from 100 to 1500 {\bf days} to get a series of $A$, $B$, $C$ and $D$ together with their uncertainties $\sigma_{A}$, $\sigma_{B}$, $\sigma_{C}$ and $\sigma_{D}$. An anti-correlation parameter $\Xi(P)$ is calculated as
\begin{equation} \label{eq_xi}
\Xi(P) = - \left( \frac{A_{1} A_{2}}{{\sigma}_{A_{1}} {\sigma}_{A_{2}}} + \frac{B_{1} B_{2}}{{\sigma}_{B_{1}} {\sigma}_{B_{2}}} \right)
\end{equation}
where the subscripts ``1" and ``2" represent the two planets. The maximum value $\Xi_{\rm max}$ represents the strongest anti-correlation among all the test periods. We select out candidate pairs with strong and significant anti-correlations by ranking their $\Xi_{\rm{max}}$.

FAPs are checked for these candidate pairs. For each pair, we scramble their TTVs randomly and obtain a similar $\Xi_{\rm max} '$. We use a superscript of `` $ \acute{} $ '' to distinguish it from $\Xi_{\rm max}$ which is obtained from the nominal data. The same process repeats $10^4$ times. The proportion of $\Xi_{\rm max} '$ larger than $\Xi_{\rm max}$ represents the {\bf expected} FAP. {\bf Following the approach of \citet{ttvs3}}, only candidate pairs with $\rm{FAP} <  10^{-3}$ are accepted \citep{ttvs3}.
	
	
\subsection{Mass Upper Limit }

Masses and free eccentricities of planet pairs near (but not in) first-order MMRs can be estimated via TTV amplitudes \citep{lithwick12,xie12}. We will estimate the mass upper limits of the accepted candidate pairs theoretically instead of simulating their stability as done by the Kepler group.

For a system near $j:j-1$ MMR, The amplitudes of a planet pair are \citep{lithwick12}:
\begin{eqnarray}
\vert V_1 \vert & = & P_1 \mu_2 \left| \frac{f}{\Delta } \right| \frac{1}{\pi j^{2/3} (j-1)^{1/3} } \label{eq_v1} \\
\vert V_2\vert & = & P_2 \mu_1 \left| \frac{g}{\Delta } \right| \frac{1}{\pi j } \label{eq_v2}
\end{eqnarray}
where $V$ is the TTV amplitude, $P$ is the orbital period, $\mu$ is the ratio of the candidate to the star, subscripts ``1'' and ``2'' represent the inner and outer candidate, g and f are Laplace coefficients.

In equation (\ref{eq_v1}) and equation (\ref{eq_v2}) we have set $| Z_{free} | = 0$ to get the mass upper limit. $Z_{free}$ is a parameter related to the complex eccentricities of the two planets. We use the model parameters corresponding to $\Xi_{\rm max}$ to calculate TTV amplitudes. Resonance order can be obtained on the basis of the ratio of the periods.

	Before we calculate the mass limit, we check the anti-correlated period $P_{\rm anti}$ and the theoretical synodic period $P_{\rm syn}$ for each pair. $P_{\rm anti}$ is the test period when $\Xi=\Xi_{\rm max}$. $P_{\rm syn}$ for a system near $j:j-1$ MMR is defined as
\begin{equation}
	P_{\rm syn} = \frac{1}{|\frac{j}{P_2}-\frac{j-1}{P_1}|}
\end{equation}

For a multiple system with only two candidates, $P_{\rm anti}$ should be equal to $P_{\rm syn}$ under ideal conditions. However, for systems containing more than two candidates or observed TTVs with large errors, the two periods will not be exactly equal. Especially for long-period candidates with {\bf a limited number of transit measurements}, the $P_{\rm anti}$ will be less accurate. When we compute mass upper limits we will use $P_{\rm syn}$ instead of $P_{\rm anti}$ to avoid its large error.

\section{CONFIRMATION OF 8 PLANETS IN 4 SYSTEMS}
We confirm 8 planets in 4 KOI systems. They are all near first-order MMRs, including one pair near 2:1 MMR (KOI-2672) and three pairs near 3:2 MMR (KOI-1236, KOI-1563 and KOI-2038). We compute $P_{\rm syn}$ of the four planet pairs. The results are listed in Table \ref{fap}. As mentioned in section 2.2,  we can see that KOI-1563 and KOI-2672 have planet pairs with almost the same $P_{\rm anti}$ and $P_{\rm syn}$, while KOI-1236 and KOI-2038 do not. Their $P_{\rm anti}$ and $P_{\rm syn}$ will be analyzed {\bf by combing} frequency spectrums in the following section. KOI-2038 was also {\bf confirmed} independently by Xie (private communication). We will describe these planet pairs individually.

\subsection{KOI-1236.01 and 1236.03}
KOI-1236 has three candidates. Moving around a star of radius 1.27 ${\rm R}_\odot$. The radii of the three candidates are $4.30\pm1.80$, $2.60\pm1.10$ and $3.10\pm1.30$ ${\rm R}_\oplus$ respectively. The periods of KOI-1236.01 and KOI-1236.03 are $35.74113$ and $54.3995$ days near 3:2 MMR. The inner most KOI-1236.02 have a period of $12.309717$ days.

\begin{figure}
\epsscale{1.0}
\plotone{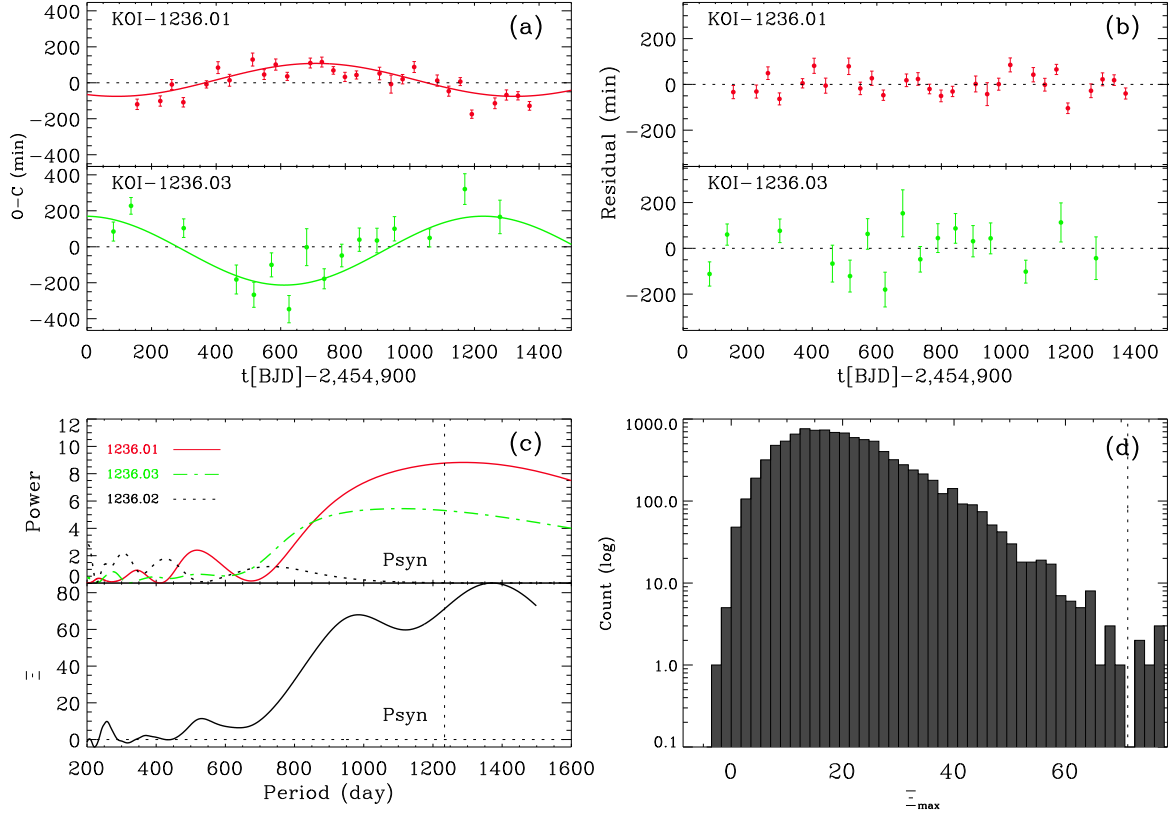}
\caption{ {\bf(a)} TTVs and the best-fit results for KOI-1236.01 and KOI-1236.03. Points with errorbars show TTVs. Solid lines correspond to the best-fit models using their theoretical synodic period ($P_{\rm syn}=1234.0$ d). {\bf(b)} Residuals between the TTVs and the best-fit results for KOI-1236.01 and KOI-1236.03. {\bf(c)} The frequency spectrums for all candidates in KOI-1236 (upper panel) and the anti-correlation curve with different test periods (lower panel). The vertical dashed line corresponds to $P_{\rm syn}$. {\bf(d)} The Monte Carlo test result of FAP. The vertical dashed line corresponds to $\Xi_{\rm max}$ calculated using $P_{\rm syn}$. The FAP for KOI-1236.01 and KOI-1236.03 is 0.0006.
\label{koi1236}}
\end{figure}

Figure \ref{koi1236}(a) and Figure \ref{koi1236}(d) illustrates distinct anti-correlation between KOI-1236.01 and 1236.03 with a low FAP=0.0006 (see Table \ref{fap}). However their $P_{\rm syn}$ is unequal to $P_{\rm anti}$ as shown in Figure \ref{koi1236}(c).  Combining the frequency spectrums of all the candidates in this system, we can see that only KOI-1236.01 and KOI-1236.03 agree with the theoretical synodic periods $P_{\rm syn}\sim1234$ d well, while KOI-1236.02 has a much smaller {\bf effect}. Therefore, it can be inferred that KOI-1236.01 and KOI-1236.03 interact with each other. The perturbations from the innermost KOI-1236.02 are very limited on both KOI-1236.01 and 1236.03. Fortunately, although the synodic period of KOI-1236.01 and KOI-1236.03 is as long as $\sim 1234$ d, it is still less than our observational duration.

We fit the TTV amplitudes of KOI-1236.01 and KOI-1236.03 with the theoretical $P_{\rm syn}$ via equation (\ref{eq_f}) to obtain their mass upper limits. Amplitudes of 91.585 min for KOI-1236.01 and 191.462 min for KOI-1236.03 are fitted here. Their residuals after TTV fitting are $19.245$ min and $44.983$ min. The residuals might be caused by the perturbations of KOI-1236.02. According to equation (\ref{eq_v1}) and (\ref{eq_v2}), we obtain the mass upper limits of $62\pm14\,{\rm M}_{\oplus}$ for KOI-1236.01 and $49\pm10\,{\rm M}_{\oplus}$ for KOI-1236.03.

\subsection{KOI-1563.01 and 1563.02}
Four candidates are moving around star KOI-1563 of radius 0.87 ${\rm R}_{\odot}$. Some properties of the host star are listed in Table \ref{star_info}. The radii of the four candidates are $3.60\pm1.30$, $3.30\pm1.10$, $2.16\pm0.77$ and $3.70\pm1.30$ ${\rm R}_{\oplus}$ respectively. The periods of KOI-1563.01 and KOI-1563.02 are 5.487006 and 8.29113 days, which are near 3:2 MMR. Other candidates have periods of 3.205322 days for KOI-1563.03 and 16.73826 days for KOI-1563.04.

\begin{figure}
\epsscale{1.0}
\plotone{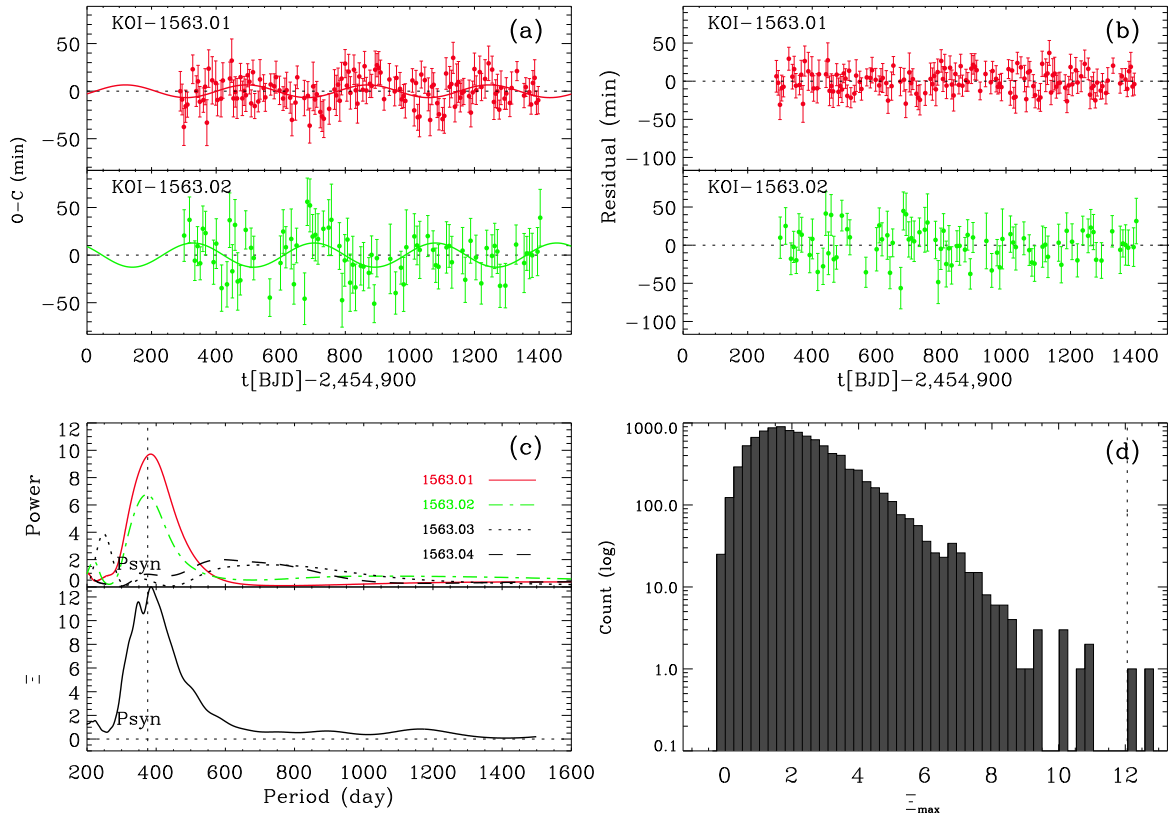}
\caption{ Similar as Figure \ref{koi1236} but for KOI-1563.01 and KOI-1563.02. Their $P_{\rm syn}=375.2$ d and FAP is 0.0002.
 \label{koi1563}}
\end{figure}

Figure \ref{koi1563}(a) and Figure \ref{koi1563}(d) shows that KOI-1563.01 and KOI-1563.02 have significant anti-correlation with a low FAP=0.0002 (see Table \ref{fap}). Their $P_{\rm syn}$ is approximately equal to $P_{\rm anti}$ as shown in Figure \ref{koi1563}(c). Therefore, it can be inferred that the perturbations of KOI-1563.03 and KOI-1563.04 on both KOI-1563.01 and KOI-1563.02 are small.

To obtain the mass upper limits of these two planets, we fit the TTV amplitudes of KOI-1563.01 and KOI-1563.02 with theoretical $P_{\rm syn}$ via equation (\ref{eq_f}). We have almost three complete synodic cycles during the observation time. Therefore we can obtain a very accurate result. An amplitude of 6.675 min for KOI-1563.01 and 12.704 min for KOI-1563.02 are fitted here. The corresponding residuals are 3.604 min and 7.608 min. Since KOI-1563.02 and 1563.04 are near 2:1 MMR  with a small $\Delta = 0.009$,  we can infer that the mass of the outermost candidate is less than $7.6\,{\rm M}_\oplus$. According to equation (3) and (4), we obtain the mass upper limits of $9.0\pm5.4\,{\rm M}_{\oplus}$ for KOI-1563.01 and $7.7\pm4.1\,M_{\oplus}$ for KOI-1563.02.

\subsection{KOI-2038.01 and 2038.02}
Star KOI-2038 of radius 0.84$R_\odot$ has four candidates. Some properties of this star are listed in Table \ref{star_info}. The radii of the candidates are $1.99\pm0.86$, $2.20\pm0.95$, $1.56\pm0.68$ and $1.61\pm0.70$ ${\rm R}_{\oplus}$ respectively. The periods of KOI-2038.01 and KOI-2038.02 are 8.305992 and 12.51217 days, which are near 3:2 MMR. Other candidates have periods of 17.91304 days for KOI-2038.03 and 25.21767 days for KOI-2038.04.

\begin{figure}
\epsscale{1.0}
\plotone{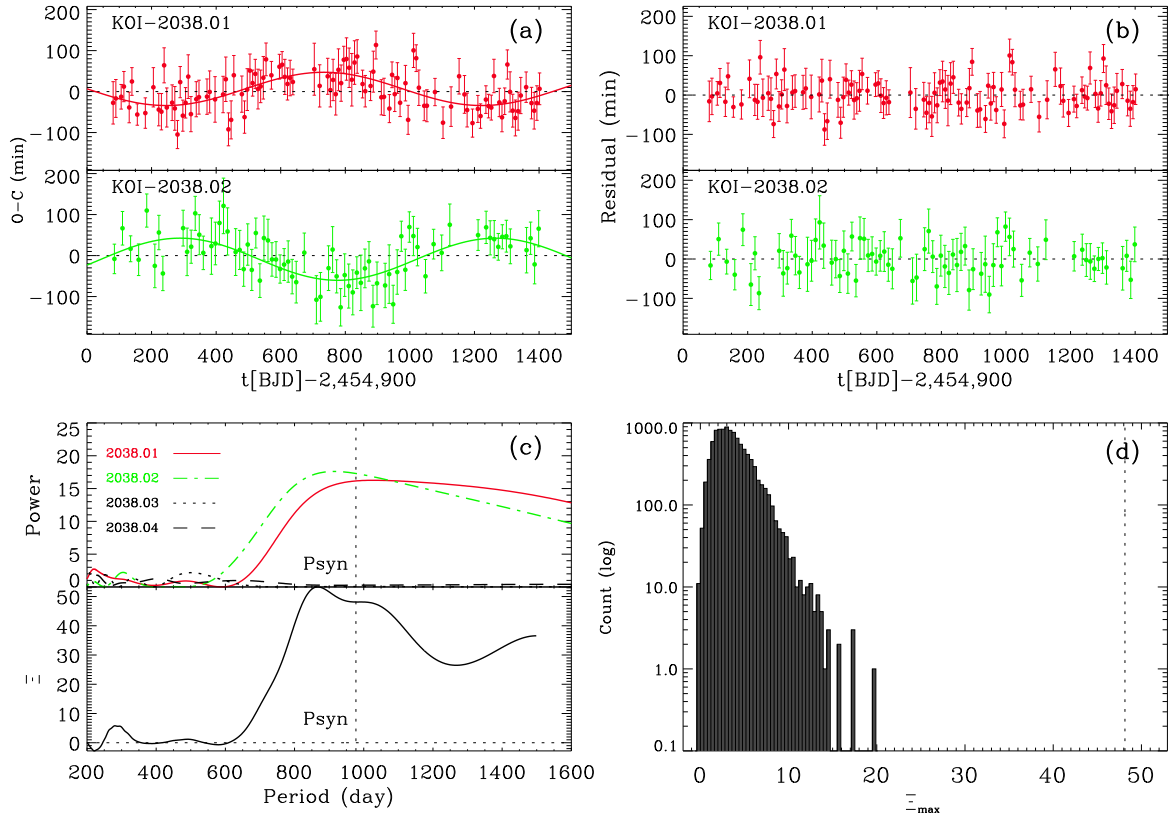}
\caption{ Similar as Figure \ref{koi1236} but for KOI-2038.01 and KOI-2038.02. Their $P_{\rm syn}=977.1$ d and  FAP is less than $10^{-4}$.
 \label{koi2038}}
\end{figure}

Figure \ref{koi2038}(a) and Figure \ref{koi2038}(d) shows that KOI-2038.01 and KOI-2038.02 have significant anti-correlation with a low FAP$< 10^{-4}$ (see Table \ref{fap}). However their $P_{\rm syn}$ is unequal to $P_{\rm anti}$ as shown in Figure \ref{koi2038}(c). Combining the frequency spectrums of all the candidates in this system, we can see only KOI-2038.01 and KOI-2038.02 agree with $P_{\rm syn} \sim 977$ d the most while others have much smaller powers. Therefore, it can be inferred that KOI-2038.01 and KOI-2038.02 interact with each other obviously. The perturbations of KOI-2038.03 and KOI-2038.04 are very limited on both of them.

To obtain the mass upper limits of these two planets, we fit the TTV amplitudes of KOI-2038.01 and KOI-2038.02 with the theoretical $P_{\rm syn}$ via equation (\ref{eq_f}). An amplitude of 40.541 min for KOI-2038.01 and 51.382 min for KOI-2038.02 are fitted here. The corresponding residuals are 11.206 min and 14.850 min. The perturbations of KOI-2038.03 and 2038.04 may be the main reason for the residuals. Assuming the residuals of KOI-2038.02 are mainly caused by KOI-2038.03, the mass of KOI-2038.03 must be less than $42\,{\rm M}_{\oplus}$. Otherwise, if the residuals of KOI-2038.02 are mainly caused by KOI-2038.04, the mass of KOI-2038.04 must be less than $8.6\,{\rm M}_{\oplus}$. According to equation (\ref{eq_v1}) and (\ref{eq_v2}), we obtain the mass upper limits of $15\pm4.3\,{\rm M}_{\oplus}$ for KOI-2038.01 and $19\pm5.2\,{\rm M}_{\oplus}$ for KOI-2038.02.

\subsection{KOI-2672.01 and 2672.02}
KOI-2672 of radius 1.04 ${\rm R}_{\odot}$, has only two candidates around. Other parameters of KOI-2672 are listed in Table \ref{star_info}. The radii of the two candidates are $5.30\pm2.10$ and $3.50\pm1.40$ ${\rm R}_{\oplus}$ respectively. The periods of KOI-2672.01 and KOI-2672.02 are 88.51658 and 42.99066 days, which are near 3:2 MMR.

\begin{figure}
\epsscale{1.0}
\plotone{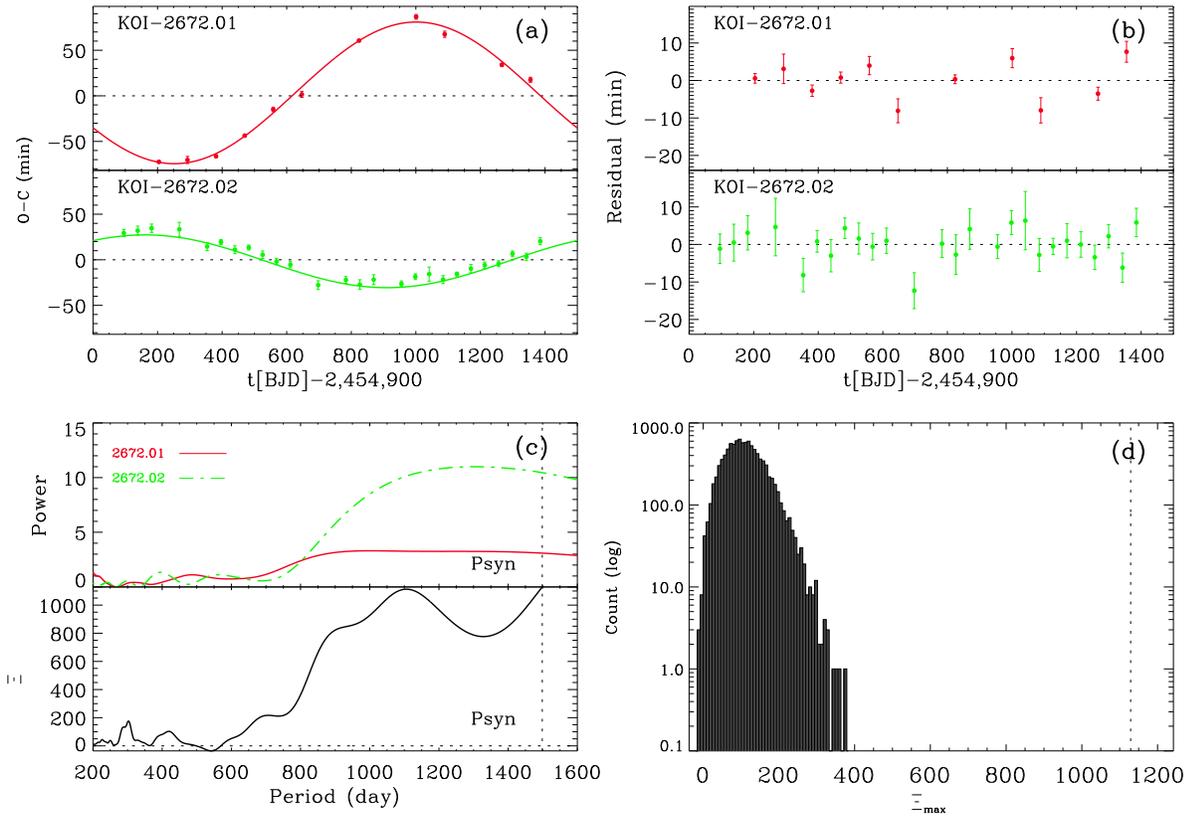}
\caption{ Similar as Figure \ref{koi1236} but for KOI-2672.01 and KOI-2672.02. Their $P_{\rm syn}=1501.0$ d and  FAP is less than $10^{-4}$.
 \label{koi2672}}
\end{figure}

Figure \ref{koi2672}(a) and Figure \ref{koi2672}(d) shows that KOI-2672.01 and KOI-2672.02 have significant anti-correlation with a low FAP$< 10^{-4}$ (see Table \ref{fap}). Their $P_{\rm syn}$ is approximately equal to $P_{\rm anti}$ as shown in Figure \ref{koi2672}(c). The strongest powers in their frequency spectrums also  coincide with $P_{\rm syn}$. Although the synodic period is as long as $\sim$ 1500 d, our observational duration still covers almost a complete cycle so we proceed with our analysis.

To obtain the mass upper limits of KOI-2672.01 and 2672.02, we fit their TTV amplitudes with theoretical $P_{\rm syn}$ via equation (\ref{eq_f}). An amplitude of 77.687 min for KOI-2672.01 and 28.952 min for KOI-2672.02 are fitted here. According to equation (\ref{eq_v1}) and (\ref{eq_v2}), we obtain the mass upper limits of $17\pm1.8\,{\rm M}_{\oplus}$ for KOI-2672.01 and $80\pm3.5\,{\rm M}_{\oplus}$ for KOI-2672.02. {\bf After removing our best-fit results via equation (\ref{eq_f}), the remaining residuals have a scatter of about 3.394 min for KOI-2672.01 and 3.115 min for KOI-2672.02.} The small residuals indicate that there is no large perturbations in this system. Assuming another planet outside is near 2:1 MMR with KOI-2672.01, we infer that the mass of the assumed planet must be less than $1.7\,{\rm M}_{\oplus}$ from equation (\ref{eq_v1}).

\section{MASS CONSTRAINT OF PLANET PAIRS}

We scan the masses of the 4 planet pairs and simulate their TTVs. For each inner planet, we set the scanning range of $M_{\rm in}$ from 0.1 to 1.5 times of the mass upper limit. For a given $M_{\rm in}$, we can obtain the mass of the outer planet $M_{\rm out0}$ via equation (\ref{eq_v1}) and (\ref{eq_v2}). Then we set the scanning range of $M_{\rm out}$ from 0.5 to 1.5 times of $M_{\rm out0}$. The large scanning ranges adopted for $M_{\rm in}$ and $M_{\rm out}$ are large enough to tolerate observational uncertainties.
The mean anomaly of the inner planet is fixed at 0, while the mean anomaly of the outer planet is changed from 0 to 350 degrees with a step of 10 degrees. The eccentricities, the argument of pericenter $\omega$ and the longitude of ascending node $\Omega$ are all fixed at 0.

\begin{figure}
\epsscale{1.0}
\plotone{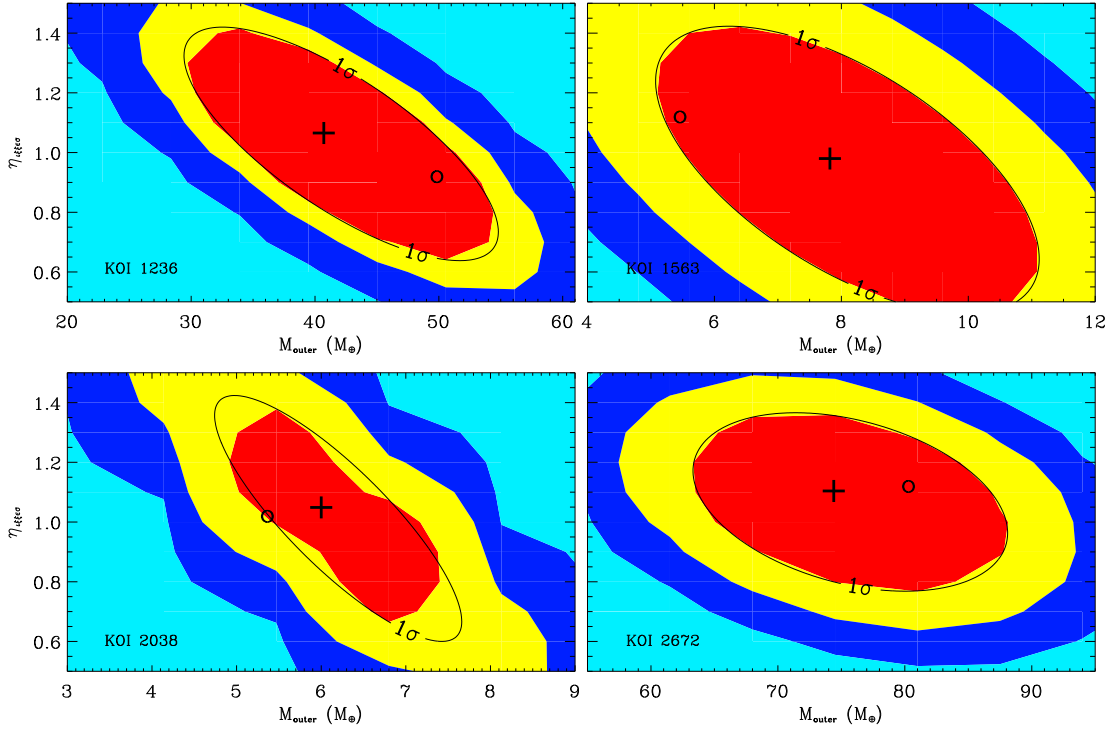}
\caption{The fraction of simulated TTVs in different mass regions with $\chi^2<\chi^2_{\rm min}+\Delta \chi^2$. $M_{\rm in}$ is the scanning mass of the inner planet. $\eta_{\rm out}$ is the ratio of the scanning mass and $M_{out0}$ (defined in Section 4). Here the minimum chi-squares $\chi^2_{\rm min}=107.3,180.1,157.3$ and $61.6$, and the degrees of freedom are 40,217,180 and 33 for KOI-1236, 1563, 2038 and 2672, respectively. We adopt $\Delta \chi^2$ = 8.2, 3.5, 5.4 and 47.6.  The plus represents the Gaussian-fit masses $M_{\rm fit}$. The circle represents the minimum chi-squared masses $M_{\rm cmin}$. The black ellipse shows the uncertainties of the Gaussian-fit masses in 1-$\sigma_{\rm fit}$. The regions in red, yellow and blue show the uncertainties of the simulated masses less than 1-$\sigma_{\rm fit}$, 2-$\sigma_{\rm fit}$ and 3-$\sigma_{\rm fit}$, respectively. The differences between $M_{\rm chi}$ and $M_{\rm fit}$ are all less than 1-$\sigma_{\rm fit}$ level.
\label{mass_fit}}
\end{figure}

\begin{figure}
\epsscale{1.0}
\plotone{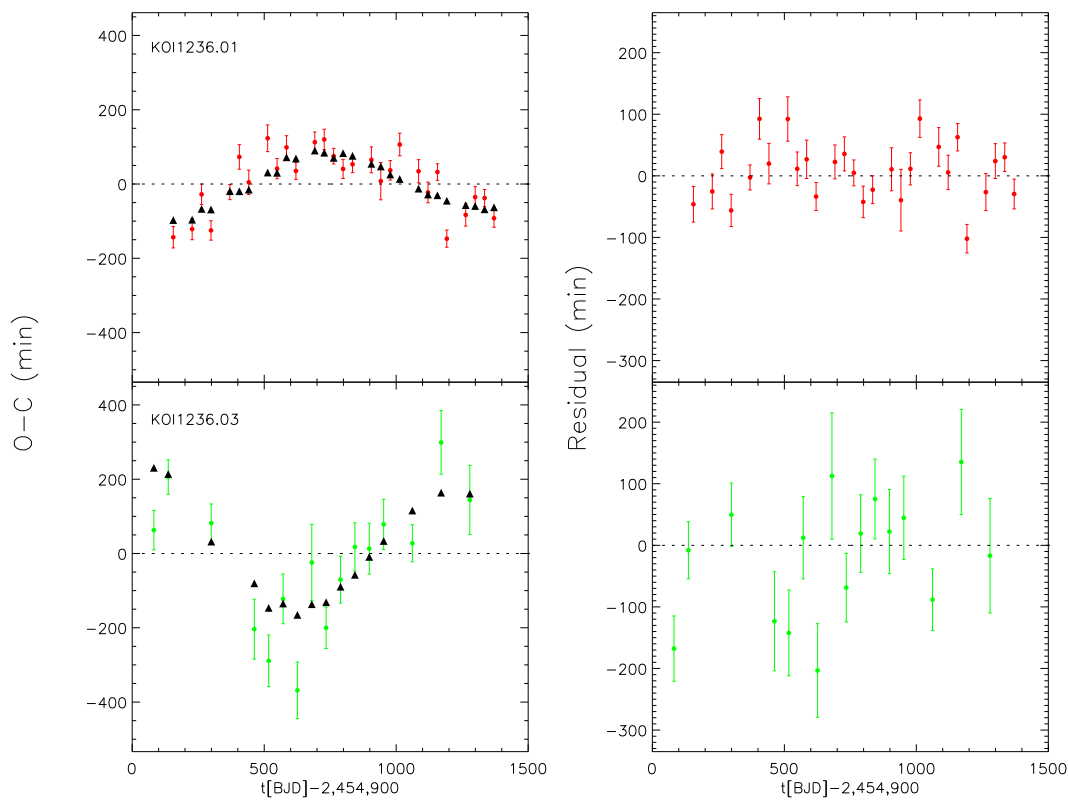}
\caption{The upper left panel shows the observed TTV (red circles) and the simulated TTV (black triangles) of KOI-1236.01.The upper right panel shows the residuals. The two panels in the bottom show observed TTV (green circles) and simulated TTV (black triangles) of KOI-1236.02,as well as residuals. Adopting minimum chi-squared masses for KOI-1236.01 and KOI-1236.03, we obtain $\chi^2_{\rm min}=107.3$ with 40 degrees of freedom.
\label{koi1236_fit}}
\end{figure}

\begin{figure}
\epsscale{1.0}
\plotone{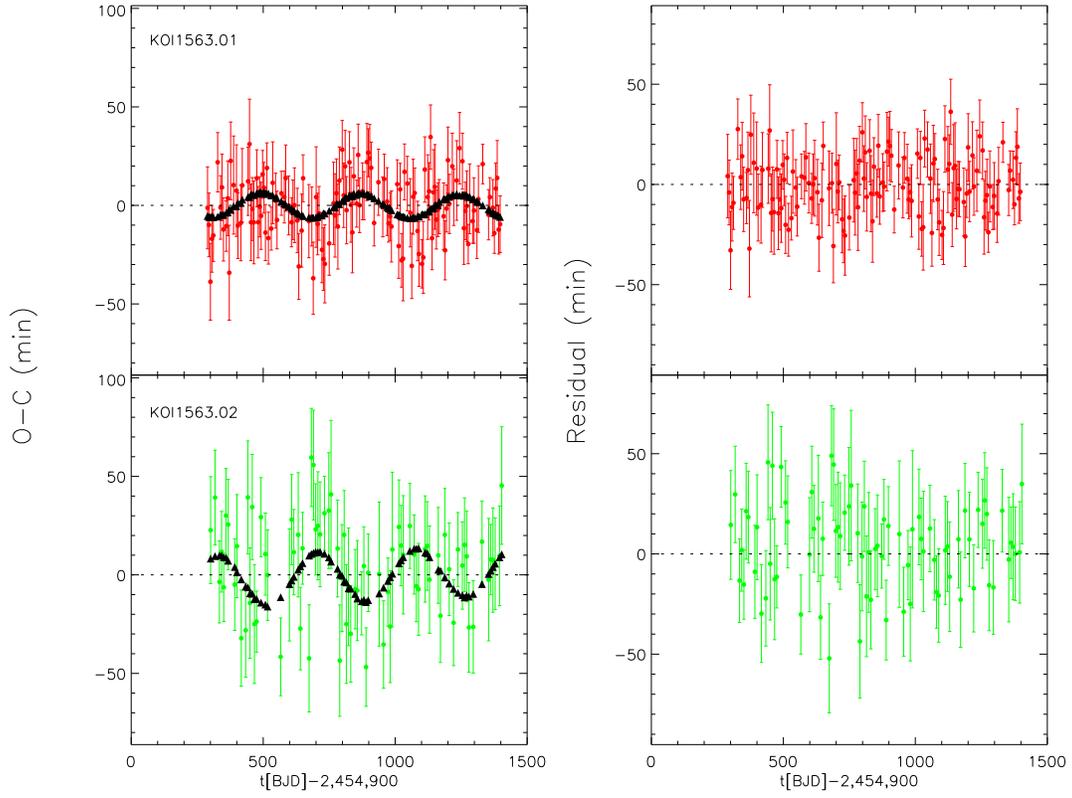}
\caption{ Similar as Figure \ref{koi1236_fit} but for KOI-1563.01 and KOI-1563.02. Our simulation gives $\chi^2_{\rm min}=180.1$ with 217 degrees of freedom.
 \label{koi1563_fit}}
\end{figure}

\begin{figure}
\epsscale{1.0}
\plotone{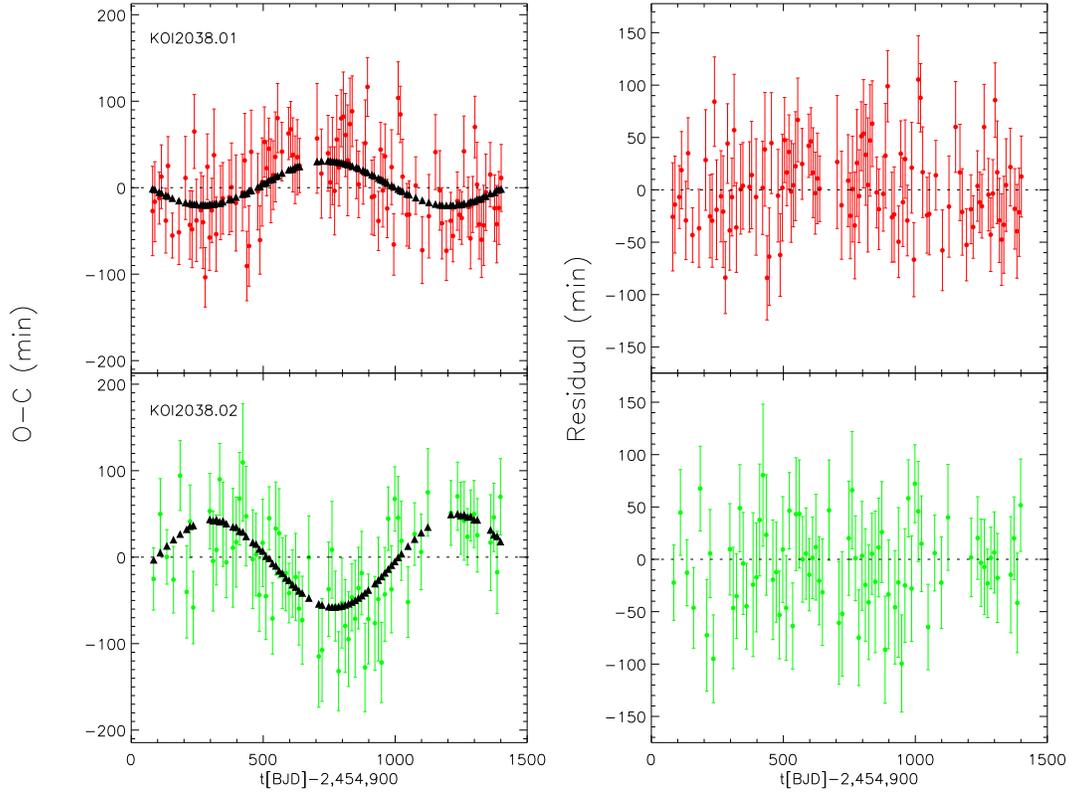}
\caption{ Similar as Figure \ref{koi1236_fit} but for KOI-2038.01 and KOI-2038.02. Our simulation gives $\chi^2_{\rm min}=157.3$ with 180 degrees  of freedom.
 \label{koi2038_fit}}
\end{figure}

\begin{figure}
\epsscale{1.0}
\plotone{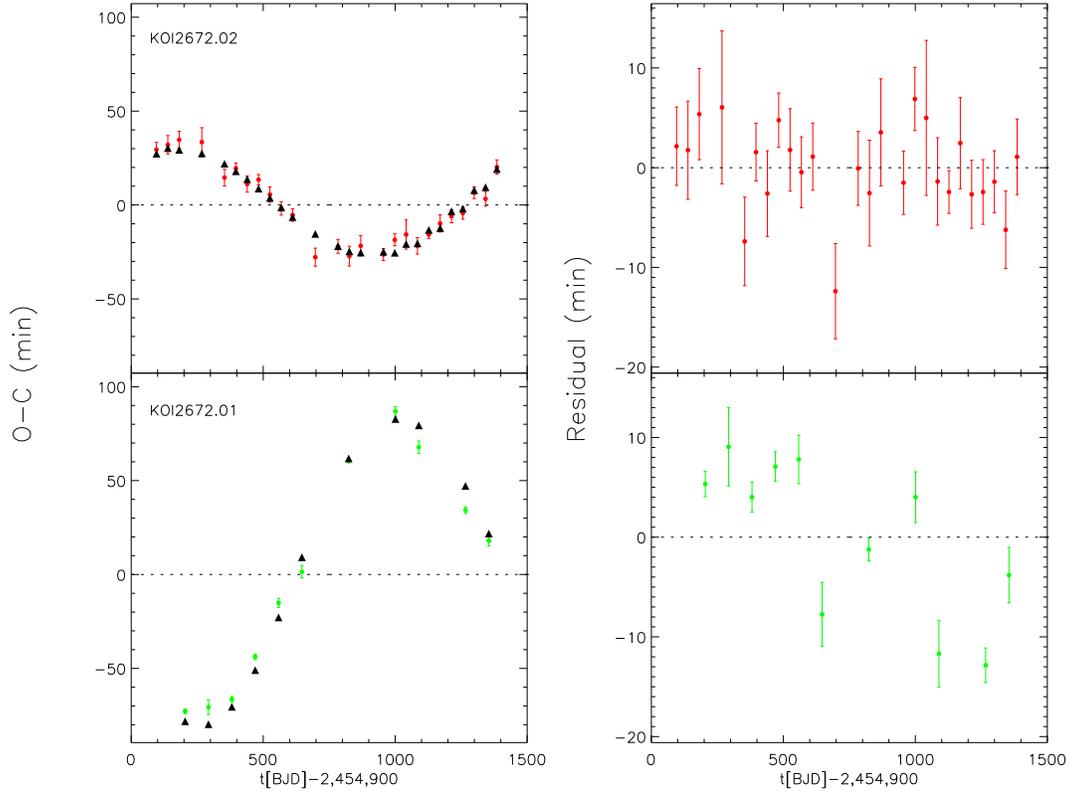}
\caption{ Similar as Figure \ref{koi1236_fit} but for KOI-2672.01 and KOI-2672.02. Our simulation gives $\chi^2_{\rm min}=61.6$ with 33 degrees of freedom.
\label{koi2672_fit}}
\end{figure}

Finally, we obtain a series of chi-squares $\chi^2$ by comparing the simulated TTVs with the observed TTVs. We can easily find the minimum chi-square $\chi^2_{\rm min}$ and {\bf the mass associated with the minimum chi-square for each planet} $M_{\rm cmin}$. For each pair of $M_{\rm in}$ and $M_{\rm out}$, we have 36 cases because we changed the mean anomaly from 0 to 350 degrees. We {\bf consider} the fraction of simulations with $\chi^2 < \chi^2_{\rm min}+\Delta \chi^2$. The values of $\chi^2_{\rm min}$ and $\Delta \chi^2$ in each system are listed in the caption of Figure \ref{mass_fit}. By applying a 2-D Gaussian fitting to the fractions, we can obtain Gaussian-fit masses $M_{\rm fit}$ and the standard deviations $\sigma_{\rm fit}$. Since $M_{\rm fit}$ is the mathematical expectation of all the simulated masses with small chi-squares, we adopt it as the most-likely mass of planet. As shown in Figure \ref{mass_fit}, the differences between $M_{\rm cmin}$ and $M_{\rm fit}$ are all less than 1-$\sigma_{\rm fit}$. Figure \ref{koi1236_fit}-\ref{koi2672_fit} {\bf show} the minimum chi-squared fit for each system.


Using the Gaussian-fit masses, we can estimate the density of each planet as shown in Table \ref{planet_info}. The densities of KOI-1236.01 and KOI-1236.02 are similar to rock planets. KOI-1563.01 and KOI-1563.02 seem to be sub-Neptunes according to their radius. However, the densities of the two planets are relatively low, and therefore they might contain a lot of gas or some other light materials. KOI-2038.01 and KOI-2038.02 are sub-Neptunes, and their densities are similar to Earth. KOI-2672.01 and KOI-2672.02 show a large density contrast.

We also check the stabilities of the four two-planet systems. According to \citet{KI02} and \citet{Zhou07}, the stable timescale depends on the separation between planets scaled by mutual hill radius. We calculate the separation of planets $\Delta a=|a_1-a_2|$ in each system, and obtain a scaled separation $k=\Delta a/R_{\rm Hill}$ , where the mutual Hill radius $R_{\rm Hill}=(\frac{\mu_1+\mu_2}{3})^{1/3}\frac{a_1+a_2}{2}$, $\mu_1$ and $\mu_2$ are the mass ratios between the planets and the host star. $a_1$ and $a_2$ are the semi-major axes of the two planets. Using the $M_{\rm fit}$ obtained above, $k$=7.2, 10.4, 10.9 and 9.8 for KOI-1236, 1563, 2038 and 2672 respectively. Obviously, the scaled separation $k>3.5$ for all four systems, which means they are stable in circular coplanar 3-body systems \citep{Glad93}. Using mercury integration package, we test the stability of each planet pairs with the best fit masses. The hybrid symplectic algorithm is adopted here. All the four systems are stable in at least 10 Myr.

\section{COMPARISON WITH CONFIRMED KEPLER PLANETS NEAR MMRS}
\begin{figure}
\epsscale{1.0}
\plotone{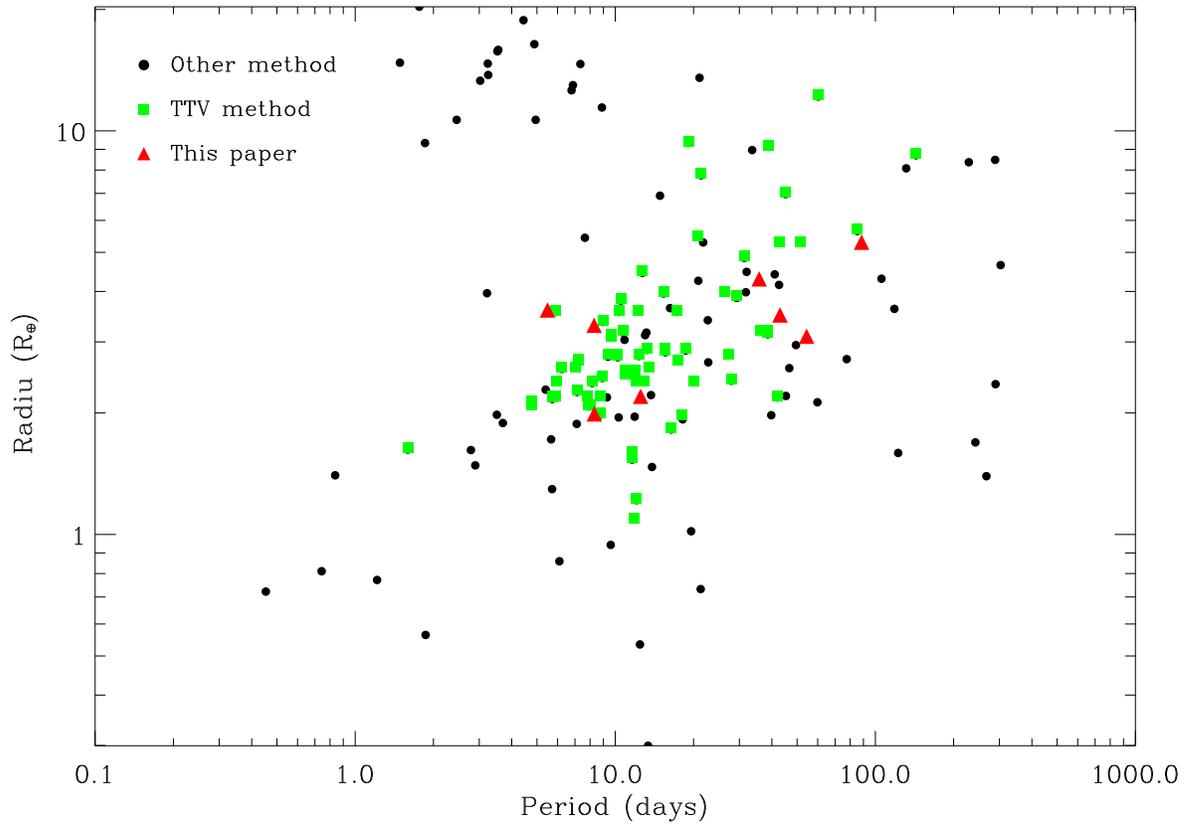}
\caption{ Period vs. radius of {\it Kepler} planets. Black filled circles are confirmed via RV or BLEND method. Green squares are confirmed via TTV method. Red triangles are confirmed in this paper, including four planets with relatively long periods ($>$ 35 day).
\label{pr_dis}}
\end{figure}

We have confirmed 8 planets via TTVs and list the parameters of them in Table \ref{planet_info}. We improve the fitting function of TTVs \citet{ttvs7} by adding a secular term. We have found a planet with a period of $88.51658$ days, which is the second longest periodical planet confirmed via TTVs under Neptune size, as shown in Figure \ref{pr_dis}. The longest confirmed via TTVs is Kepler-30 d, which is near 5:2 MMR with Kepler-30 c \citep{ttvs4}. Although Kepler-30 d has only five TTV data, it can also be confirmed by analyzing the TTVs of 30 b and 30 c, using a transiting starspot model \citep{SO12}. However, KOI-2672.01 confirmed in this paper is presently the longest periodical planet near first-order MMR. In addition, this pair also has a large density contrast. A similar system is Kepler-36 \citep{carter12}. {\bf As pointed by \citet{lopez13}, the large density contrast of Kepler-36 can be interpreted as a result of photo evaporation of the inner planet. However, this interpretation may not suitable for KOI-2672 system. The inner planet KOI-2672.02 is much massive than a normal core mass and far from the host star, therefore its photo evaporation may not very strong. We infer that the less dense planet KOI-2672.01 has a small core mass and may have formed inside the ``snow line'', while KOI-2672.02 
may have formed outside and might experience obvious collisions
to become so dense and massive.} After dynamical evolution, especially migrations and collisions, they are {\bf driven into their present architecture}. We also note that this system has only two planets detected.

\begin{figure}
\epsscale{1.0}
\plotone{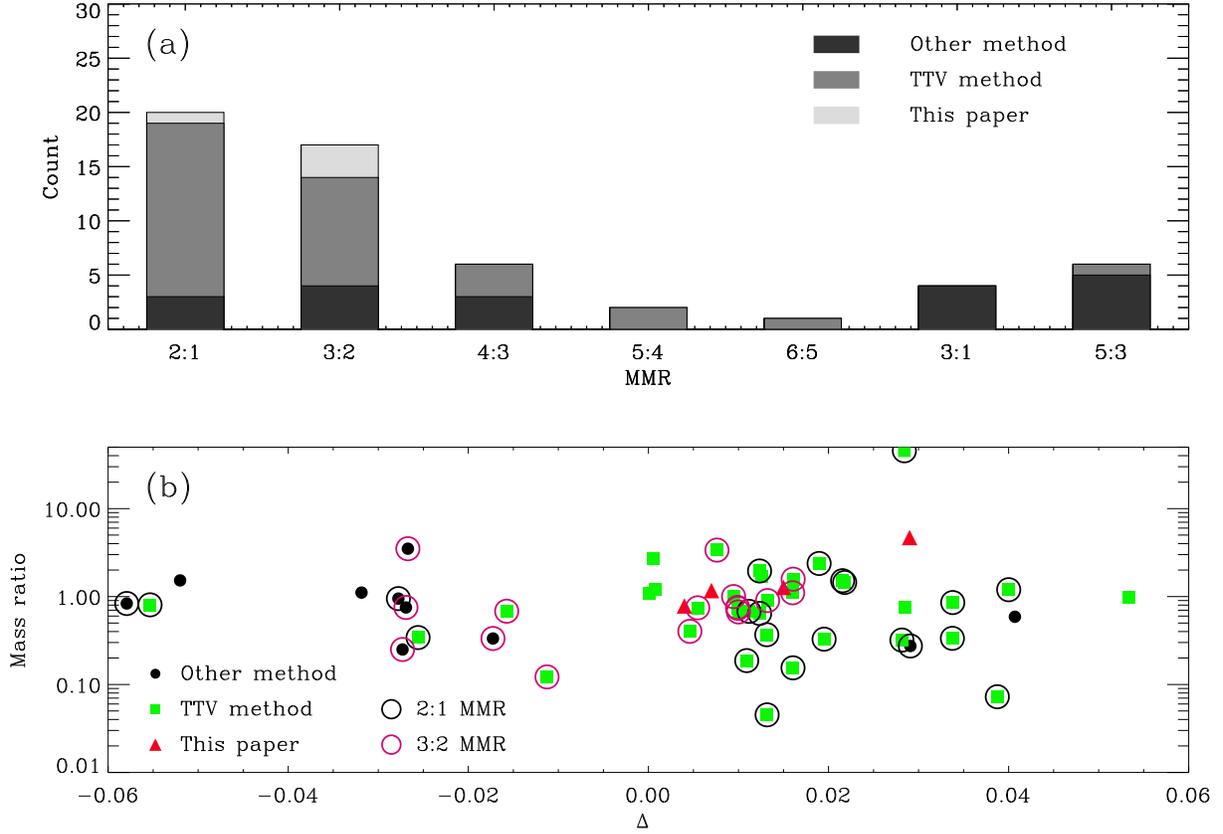}
\caption{ {\bf(a)} Distribution of the MMRs of {\it Kepler} planet pairs. Dark grey are confirmed via RV or BLEND method. Grey are confirmed via TTV method. Light grey are confirmed in this paper. In this paper we add one pair near 2:1 MMR and 3 pairs near 3:2 MMR. {\bf(b)} Parameter $\Delta$ (see equation(\ref{eq_delta})) vs. mass ratio in first-order MMRs of {\it Kepler} planets. $\Delta$ of most planet pairs are greater than 0, which infers that the outer planets are prefer to stay a little far away from the exact first-order MMR position \citep{fabrycky12}.
\label{mmr_hist}}
\end{figure}

We show all the Kepler planets near MMRs as shown in Figure \ref{mmr_hist} and compare our results with them. Adding one pair near 2:1 MMR and three pairs near 3:2 MMR, there are totally 20 pairs near 2:1 MMR and 17 pairs near 3:2 MMR. Only 10 pairs are in other first-order MMRs and 10 pairs in the second-order MMRs. Theoretical works show that, assuming these configurations were through type I migration of planetary embryos with mass up to super-Earth \citep{GT79,Tan02,Ward97}, the final architectures are mainly determined by the migration speed, e.g., low speed migration can stall at 2:1 MMR, while high speed pass 2:1 MMR and will stall at j:j-1 with higher j (e.g. \citealt{Zhou10,SP10,wang12}). Supposing all the planet pairs are initially located outside the 2:1 MMR, the frequent outcome of planet pairs at 2:1 and 3:2 MMR (Figure \ref{mmr_hist}) is consistent with the migration scenario. Figure \ref{mmr_hist} presents $\Delta$ as defined by equation (\ref{eq_delta}) and the mass ratio of the planet pair of each system. It's obvious that 75\% of the planet pairs have a period ratio that are a little larger than the exact value $j : (j-1)$. This is revealed by \citet{fabrycky12} and interpreted by \citet{lithwu12} and \citet{hoi13} via tidal dissipation.

\section{SUMMARY AND DISCUSSION}
We have found 4 planet pairs near first-order MMRs via TTVs with a time span as long as $\sim$1350 days, i.e., KOI-1236.01 and KOI-1236-03 near 3:2 MMR, KOI-1563.01 and KOI-1563.02 near 3:2 MMR, KOI-2038.01 and KOI-2038.02 near 3:2 MMR and KOI-2672.01 and KOI-2672.02 near 2:1 MMR. We also estimate their mass upper limits via equation (\ref{eq_v1}) and (\ref{eq_v2}), as shown in Table \ref{planet_info}. KOI-2672 has only two planets with a long theoretical $P_{\rm syn}$ about 1500 d, and the outer planet has the longest period among the planet pairs near first-order MMRs.

By simulating the TTVs of planet pairs, we show the minimum chi-squared masses and the Gaussian-fit masses of planet pairs. All the two-planet systems with the Gaussian-fit masses are stable for at least 10 Myr. We calculate the density of each planet according to their best-fit mass. Planets in KOI-1236 and KOI-2038 system seem like rock planets while planets in KOI-1563 system have a relative low density. Considering the large difference between the densities of 2672.01 and 2672.02, we infer that they formed in different regions and evolved in current orbital architecture due to migration or some other mechanisms.

We also compare our new confirmed planets with other Kepler planets, especially these planet pairs near MMRs. Adding the new four pairs near MMRs, we find that about 36\% and 30\% of the planet pairs are near 2:1 and 3:2 MMR, respectively.  A few more planet pairs are near 2:1 MMR than are near 3:2 MMR. If this trend continues, it is important of constraining the migrating rate of planets during the evolution of multi-planet systems \citep{OK13}. Planet pairs prefer to keep an architecture that the outside planets are a little outside than the exact positions with period ratios of $j:j-1$, which is consistent with \citet{fabrycky12}.

In this paper, we only consider planet pairs near first-order MMRs. Additionally, planet pairs near higher order MMRs have a weaker TTV signal than planets near first-order MMRs, thus it is much harder to estimate their mass upper limits. However, the anti-correlations of their TTVs and small FAPs are also available to confirm these gravitational pairs. The fraction of planet pairs near high order MMRs are relatively rare, only 10 pairs are confirmed by the observation from {\it Kepler}.

Here we use the TTVs spanning up to about 1350 days. TTVs in a longer timescale can provide more information about the secular influences in planet pairs near or in MMRs \citep{ketchum13,boue12}. Subsequent observations of transiting planets or {\it Kepler} candidates provide us more information about their TTVs to investigate the dynamical properties of multi-planet systems and might confirm planets with longer period in the habitable zone.

\acknowledgments
We appreciate NASA and the Kepler group for the great work on exoplanet detection. This work is supported by the Key Development Program of Basic Research of China(No. 2013CB834900), National Natural Science Foundations of China (Nos. 10925313, 11003010 and 11333002), 985 Project of Ministration of Education and Superiority Discipline Construction Project of Jiangsu Province. We are also grateful to High Performance Computing Center (HPCC) of Nanjing University for the catalog refinement process.

\clearpage

\begin{deluxetable}{llccccc}
\tablewidth{0pt}
\tablecaption{ Anti-correlations and FAPs of 8 candidates in 4 systems }
\tablehead{
\colhead{KOI} & \colhead{Anti-correlation pair} & \colhead{FAP}  & \colhead{ $P_{\rm anti}$ }   &
\colhead{$P_{\rm syn}$} & \colhead{ $P_j / P_{j-1}$ }  & \colhead{$\Delta$}                \\
\colhead{} & \colhead{} & \colhead{} & \colhead{(days)} & \colhead{(days)} & \colhead{} & \colhead{}
}
\startdata
1236	&	KOI1236.01 - KOI1236.03	&	0.0006	&	1498.7	&	1234.0	&	0.657	&	0.015	\\
1563	&	KOI1563.01 - KOI1563.02	&	0.0002	&	384.6	&	375.2	&	0.662	&	0.007	\\
2038	&	KOI2038.01 - KOI2038.02	&	$<$0.0001	&	863.0	&	977.1	&	0.664	&	0.004	\\
2672	&	KOI2672.01 - KOI2672.02	&	$<$0.0001	&	1449.3	&	1501.0	&	0.486	&	0.029	\\
\enddata
\label{fap}
\end{deluxetable}

\begin{deluxetable}{cccccccccc}
\tabletypesize{\scriptsize}
\tablewidth{0pt}
\tablecaption{Main properties of confirmed planets}
\tablehead{
\colhead{KOI}  & \colhead{ $\rm{T}_0$}  & \colhead{Period} &
\colhead{a}    &  \colhead{Teq}         & \colhead{Duration}          & \colhead{Radius}       &
\colhead{$\rm{M}_{\rm max}$}  & \colhead{$\rm{M}_{\rm fit}$}    & \colhead{Density} \\
\colhead{}     & \colhead{ (BJD-2,454,900) }      & \colhead{(days)}  &   \colhead{(AU)}                 &
\colhead{(K)} &  \colhead{(hr)}         &  \colhead{(${\rm R}_\oplus$)}  &   \colhead{(${\rm M}_\oplus$)}       &
\colhead{(${\rm M}_\oplus$)} &   \colhead{(g/$\rm{cm}^3$)} }
\startdata
1236.01	&	84.06120	&	35.741130	&	0.23238	&	699	&	8.1417	&	4.30	&	62	&	 44	&	3.05	 \\
1236.03	&	81.87990	&	54.399500	&	0.30728	&	607	&	7.7780	&	3.10	&	49	&	 31	&	5.74	 \\
1563.01	&	289.07960	&	5.487006	&	0.05856	&	833	&	2.9700	&	3.60	&	9.0	&	 8.1&	0.96	 \\
1563.02	&	292.62760	&	8.291130	&	0.07711	&	729	&	3.2008	&	3.30	&	7.7	&	 6.6&	1.01	 \\
2038.01	&	72.71660	&	8.305992	&	0.07891	&	814	&	3.8461	&	1.99	&	15	&	 6.1&	4.27	 \\
2038.02	&	72.45610	&	12.512170	&	0.10369	&	712	&	4.3714	&	2.20	&	19	&	 7.0&	3.63	 \\
2672.01	&	115.65259	&	88.516580	&	0.36709	&	413	&	6.8748	&	5.30	&	17	&	 18	&	0.67	 \\
2672.02	&	95.51176	&	42.990660	&	0.22662	&	525	&	4.7047	&	3.50	&	80	&	 75	&	9.66	 \\
\enddata
\label{planet_info}
\begin{flushleft}
Note. For each planet, Column 1 to 10 represent the KOI number, transiting offsets, period, semi-major axis, equilibrium temperature, transiting duration, planetary radius, mass upper limit, Gaussian-fit mass and density, respectively. Data in column 1-7 are from Kepler website, and data in column 8-10 are from this paper.
\end{flushleft}
\end{deluxetable}

\begin{deluxetable}{ccccccccc}
\tablewidth{0pt}
\tablecaption{ Stellar properties of confirmed multi-planet systems. }
\tablehead{
\colhead{KOI} & \colhead{ KIC} & \colhead{Kp} & \colhead{Teff} & \colhead{log(g)}  &
\colhead{R*}  & \colhead{M*}   & \colhead{RA} & \colhead{DEC}  \\
\colhead{}    & \colhead{}     & \colhead{(mag)} & \colhead{(K)}  & \colhead{(cm/$\rm{s}^2$)}   & \colhead{(${\rm R}_\odot$)}     &  \colhead{(${\rm M}_\odot$)} & \colhead{(J2000)} & \colhead{(J2000)}
}
\startdata
1236	&	6677841	&	13.659	&	6779	&	4.35	&	1.27	&	1.31	&	19 09 33.889	&	 +42 11 41.40	\\
1563	&	5219234	&	15.812	&	4918	&	4.51	&	0.87	&	0.89	&	19 56 53.840	&	 +40 20 35.46	\\
2038	&	8950568	&	14.779	&	5666	&	4.57	&	0.84	&	0.95	&	19 23 53.621	&	 +45 17 25.16	\\
2672	&	11253827	&	11.921	&	5565	&	4.33	&	1.04	&	0.84	&	19 44 31.875	 &	 +48 58 38.65	\\
\enddata
\label{star_info}
\end{deluxetable}

\clearpage

\begin{figure}
\epsscale{1.0}
\plotone{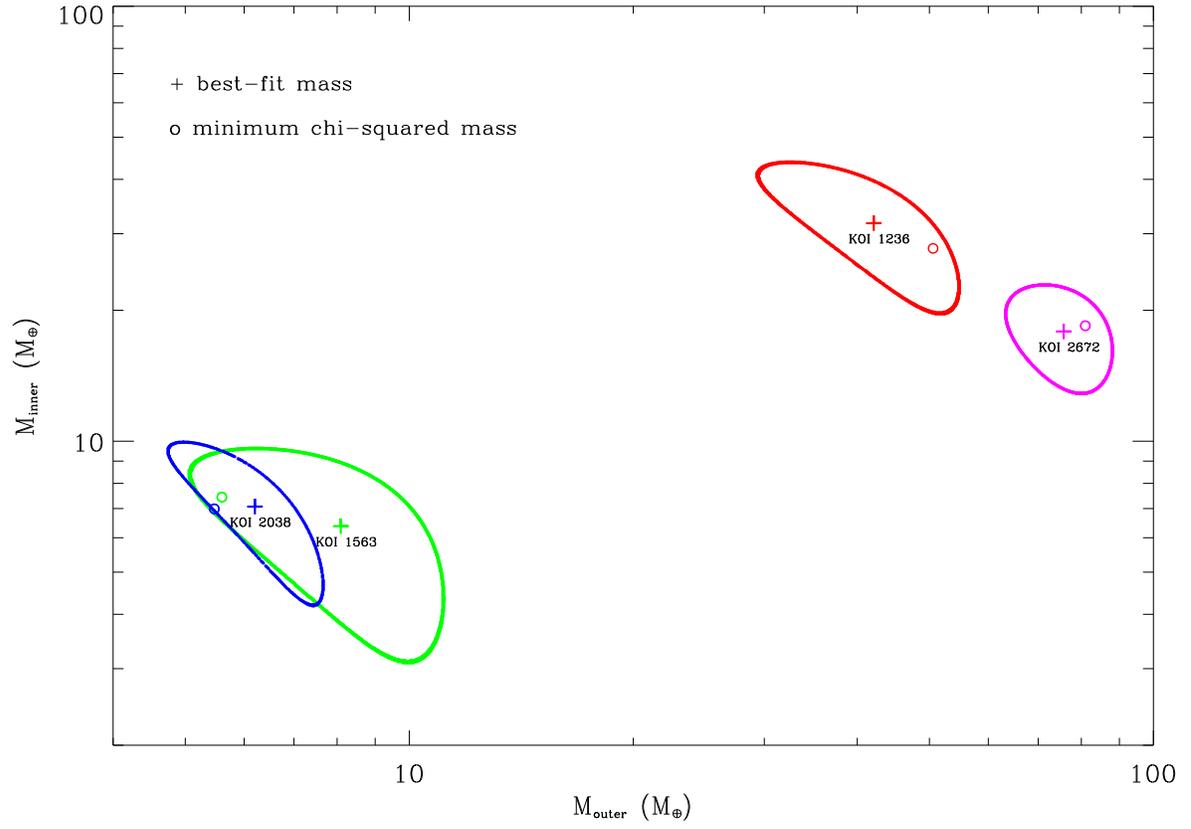}
\caption{ The Gaussian-fit masses (`+') and the minimum chi-squared masses (`o') of all planets. Colored islands show the uncertainties of the Gaussian-fit masses in 1-$\sigma_{\rm fit}$.
\label{mass_fit1sig}}
\end{figure}

\clearpage


\begin{thebibliography}{}
\bibitem[Agol et al.(2005)]{agol05} Agol, E., Steffen, J., Sari, R., \& Clarkson, W.\ 2005, \mnras, 359, 567
\bibitem[Akeson et al.(2013)]{akeson13} Akeson, R.~L., Chen, X., Ciardi, D., et al.\ 2013, arXiv:1307.2944
\bibitem[Ballard et al.(2011)]{ballard11} Ballard, S., Fabrycky, D., Fressin, F., et al.\ 2011, \apj, 743, 200
\bibitem[Batalha et al.(2013)]{Batalha13} Batalha, N.~M., Rowe, J.~F., Bryson, S.~T., et al.\ 2013, \apjs, 204, 24
\bibitem[Bou{\'e} et al.(2012)]{boue12} Bou{\'e}, G., Oshagh, M., Montalto, M., \& Santos, N.~C.\ 2012, \mnras, 422, L57
\bibitem[Borucki et al.(2010)]{borucki10} Borucki, W.~J., Koch, D., Basri, G., et al.\ 2010, Science, 327, 977
\bibitem[Borucki et al.(2011)]{borucki11} Borucki, W.~J., Koch, D.~G., Basri, G., et al.\ 2011, \apj, 736, 19
\bibitem[Carter et al.(2012)]{carter12} Carter, J.~A., Agol, E.,
Chaplin, W.~J., et al.\ 2012, Science, 337, 556
\bibitem[Fabrycky et al.(2012a)]{ttvs4} Fabrycky, D.~C., Ford, E.~B., Steffen, J.~H., et al.\ 2012, \apj, 750, 114
\bibitem[Fabrycky et al.(2012b)]{fabrycky12} Fabrycky, D.~C., Lissauer, J.~J., Ragozzine, D., et al.\ 2012, arXiv:1202.6328
\bibitem[Ford et al.(2012)]{ttvs2} Ford, E.~B., Fabrycky, D.~C., Steffen, J.~H., et al.\ 2012, \apj, 750, 113
\bibitem[Gladman(1993)]{Glad93}Gladman, B., 1993, Icarus, 106, 247
\bibitem[Goldreich \& Tremaine(1979)]{GT79}Goldreich, P., \& Tremaine, S., 1979, ApJ, 233, 857
\bibitem[Holman \& Murray(2005)]{hm05} Holman, M.~J., \& Murray, N.~W.\ 2005, Science, 307, 1288
\bibitem[Holman et al.(2010)]{holman10} Holman, M.~J., Fabrycky, D.~C., Ragozzine, D., et al.\ 2010, Science, 330, 51
\bibitem[Ketchum et al.(2013)]{ketchum13} Ketchum, J.~A., Adams, F.~C., \& Bloch, A.~M.\ 2013, \apj, 762, 71
\bibitem[Koch et al.(2010)]{koch10} Koch, D.~G., Borucki, W.~J., Basri, G., et al.\ 2010, \apjl, 713, L79
\bibitem[Kokubo \& Ida(2002)]{KI02} Kokubo, E. \& Ida, S., 2002, ApJ, 581, 666
\bibitem[Lee et al.(2013)]{hoi13} Lee, M.~H., Fabrycky, D., \& Lin, D.~N.~C.\ 2013, arXiv:1307.4874
\bibitem[Lithwick et al.(2012)]{lithwick12} Lithwick, Y., Xie, J., \& Wu, Y.\ 2012, \apj, 761, 122
\bibitem[Lithwick \& Wu(2012)]{lithwu12} Lithwick, Y., \& Wu, Y.\ 2012, \apjl, 756, L11
\bibitem[Lopez \& Fortney(2013)]{lopez13} Lopez, E., \& Fortney, J.\ 2013, arXiv:1305.0269 
\bibitem[Mandel \& Agol(2002)]{mandel02} Mandel, K., \& Agol, E.\ 2002, \apjl, 580, L171
\bibitem[Mazeh et al.(2013)]{ttvs8} Mazeh, T., Nachmani, G., Holczer, T., et al.\ 2013, arXiv:1301.5499
\bibitem[Nesvorn{\'y} et al.(2012)]{nesvorn12} Nesvorn{\'y}, D., Kipping, D.~M., Buchhave, L.~A., et al.\ 2012, Science, 336, 1133
\bibitem[Ogihara \& Kobayashi(2013)]{OK13} Ogihara, M. \& Kobayashi, H., \ 2013, arxiv: 1307.7776
\bibitem[Tanaka et al.(2002)]{Tan02}Tanaka, H., Takeuchi, T., \& Ward, W. R. 2002, ApJ, 565, 1257
\bibitem[Ward(1997)]{Ward97}Ward, W. R. 1997, Icarus, 126, 261
\bibitem[Wang et al.(2012)]{wang12} Wang, S., Ji, J., \& Zhou, J.-L.\ 2012, \apj, 753, 170
\bibitem[Sanchis-Ojeda et al.(2012)]{SO12} Sanchis-Ojeda, R., Fabrycky, D. C., Winn, J. N. et al., 2012, Nature, 487, 449
\bibitem[Spiegel et al.(2011)]{spiegel11} Spiegel, D.~S., Burrows, A., \& Milsom, J.~A.\ 2011, \apj, 727, 57
\bibitem[Steffen et al.(2012a)]{ttvs3} Steffen, J.~H., Fabrycky, D.~C., Ford, E.~B., et al.\ 2012, \mnras, 421, 2342
\bibitem[Steffen et al.(2012b)]{ttvs6} Steffen, J.~H., Ford, E.~B., Rowe, J.~F., et al.\ 2012, \apj, 756, 186
\bibitem[Steffen et al.(2013)]{ttvs7} Steffen, J.~H., Fabrycky, D.~C., Agol, E., et al.\ 2013, \mnras, 428, 1077
\bibitem[Szuszkiewicz \& Papaloizou(2010)]{SP10} Szuszkiewicz, E. \& Papaloizou, J. C. B., 2010, EAS, 42, 303S
\bibitem[Xie(2012)]{xie12} Xie, J.-W.\ 2012, arXiv:1208.3312
\bibitem[Zhou et al.(2007)]{Zhou07} Zhou, J. L., Lin, D. N. C. \& Sun, Y. S., 2007, ApJ, 666, 423
\bibitem[Zhou(2010)]{Zhou10} Zhou J.-L., 2010, EAS, 42, 255Z

\end{thebibliography}
\end{document}